\documentclass[%
 reprint,
superscriptaddress,
showpacs,
 amsmath,amssymb,
 aps,
prl,
]{revtex4-1}

\usepackage{graphicx}
\usepackage{dcolumn}
\usepackage{bm}
\usepackage{amsmath}
\usepackage{autobreak}
\usepackage{hyperref}
\usepackage[mathlines]{lineno}
\usepackage{natbib}
\usepackage{multirow}
\usepackage{booktabs}

\begin{document}

\preprint{APS/123-QED}

\title{Constraints on the Blazar-Boosted Dark Matter from the CDEX-10 Experiment}

\author{R. Xu}
\affiliation{Key Laboratory of Particle and Radiation Imaging (Ministry of Education) and Department of Engineering Physics, Tsinghua University, Beijing 100084}
\author{L.~T.~Yang}
\email{Corresponding author: yanglt@mail.tsinghua.edu.cn}
\affiliation{Key Laboratory of Particle and Radiation Imaging (Ministry of Education) and Department of Engineering Physics, Tsinghua University, Beijing 100084}
\author{Q.~Yue}
\email{Corresponding author: yueq@mail.tsinghua.edu.cn}
\affiliation{Key Laboratory of Particle and Radiation Imaging (Ministry of Education) and Department of Engineering Physics, Tsinghua University, Beijing 100084}
\author{K.~J.~Kang}
\affiliation{Key Laboratory of Particle and Radiation Imaging (Ministry of Education) and Department of Engineering Physics, Tsinghua University, Beijing 100084}
\author{Y.~J.~Li}
\affiliation{Key Laboratory of Particle and Radiation Imaging (Ministry of Education) and Department of Engineering Physics, Tsinghua University, Beijing 100084}

\author{H.~P.~An}
\affiliation{Key Laboratory of Particle and Radiation Imaging (Ministry of Education) and Department of Engineering Physics, Tsinghua University, Beijing 100084}
\affiliation{Department of Physics, Tsinghua University, Beijing 100084}

\author{Greeshma~C.}
\altaffiliation{Participating as a member of TEXONO Collaboration}
\affiliation{Institute of Physics, Academia Sinica, Taipei 11529}

\author{J.~P.~Chang}
\affiliation{NUCTECH Company, Beijing 100084}

\author{Y.~H.~Chen}
\affiliation{YaLong River Hydropower Development Company, Chengdu 610051}
\author{J.~P.~Cheng}
\affiliation{Key Laboratory of Particle and Radiation Imaging (Ministry of Education) and Department of Engineering Physics, Tsinghua University, Beijing 100084}
\affiliation{College of Nuclear Science and Technology, Beijing Normal University, Beijing 100875}
\author{W.~H.~Dai}
\affiliation{Key Laboratory of Particle and Radiation Imaging (Ministry of Education) and Department of Engineering Physics, Tsinghua University, Beijing 100084}
\author{Z.~Deng}
\affiliation{Key Laboratory of Particle and Radiation Imaging (Ministry of Education) and Department of Engineering Physics, Tsinghua University, Beijing 100084}
\author{C.~H.~Fang}
\affiliation{College of Physics, Sichuan University, Chengdu 610065}
\author{X.~P.~Geng}
\affiliation{Key Laboratory of Particle and Radiation Imaging (Ministry of Education) and Department of Engineering Physics, Tsinghua University, Beijing 100084}
\author{H.~Gong}
\affiliation{Key Laboratory of Particle and Radiation Imaging (Ministry of Education) and Department of Engineering Physics, Tsinghua University, Beijing 100084}
\author{Q.~J.~Guo}
\affiliation{School of Physics, Peking University, Beijing 100871}
\author{T.~Guo}
\affiliation{Key Laboratory of Particle and Radiation Imaging (Ministry of Education) and Department of Engineering Physics, Tsinghua University, Beijing 100084}
\author{X.~Y.~Guo}
\affiliation{YaLong River Hydropower Development Company, Chengdu 610051}
\author{L.~He}
\affiliation{NUCTECH Company, Beijing 100084}
\author{S.~M.~He}
\affiliation{YaLong River Hydropower Development Company, Chengdu 610051}
\author{J.~W.~Hu}
\affiliation{Key Laboratory of Particle and Radiation Imaging (Ministry of Education) and Department of Engineering Physics, Tsinghua University, Beijing 100084}
\author{H.~X.~Huang}
\affiliation{Department of Nuclear Physics, China Institute of Atomic Energy, Beijing 102413}
\author{T.~C.~Huang}
\affiliation{Sino-French Institute of Nuclear and Technology, Sun Yat-sen University, Zhuhai 519082}
\author{L.~Jiang}
\affiliation{Key Laboratory of Particle and Radiation Imaging (Ministry of Education) and Department of Engineering Physics, Tsinghua University, Beijing 100084}
\author{S.~Karmakar}
\altaffiliation{Participating as a member of TEXONO Collaboration}
\affiliation{Institute of Physics, Academia Sinica, Taipei 11529}

\author{H.~B.~Li}
\altaffiliation{Participating as a member of TEXONO Collaboration}
\affiliation{Institute of Physics, Academia Sinica, Taipei 11529}
\author{H.~Y.~Li}
\affiliation{College of Physics, Sichuan University, Chengdu 610065}
\author{J.~M.~Li}
\affiliation{Key Laboratory of Particle and Radiation Imaging (Ministry of Education) and Department of Engineering Physics, Tsinghua University, Beijing 100084}
\author{J.~Li}
\affiliation{Key Laboratory of Particle and Radiation Imaging (Ministry of Education) and Department of Engineering Physics, Tsinghua University, Beijing 100084}
\author{Q.~Y.~Li}
\affiliation{College of Physics, Sichuan University, Chengdu 610065}
\author{R.~M.~J.~Li}
\affiliation{College of Physics, Sichuan University, Chengdu 610065}
\author{X.~Q.~Li}
\affiliation{School of Physics, Nankai University, Tianjin 300071}
\author{Y.~L.~Li}
\affiliation{Key Laboratory of Particle and Radiation Imaging (Ministry of Education) and Department of Engineering Physics, Tsinghua University, Beijing 100084}
\author{Y.~F.~Liang}
\affiliation{Key Laboratory of Particle and Radiation Imaging (Ministry of Education) and Department of Engineering Physics, Tsinghua University, Beijing 100084}
\author{B.~Liao}
\affiliation{College of Nuclear Science and Technology, Beijing Normal University, Beijing 100875}
\author{F.~K.~Lin}
\altaffiliation{Participating as a member of TEXONO Collaboration}
\affiliation{Institute of Physics, Academia Sinica, Taipei 11529}
\author{S.~T.~Lin}
\affiliation{College of Physics, Sichuan University, Chengdu 610065}
\author{J.~X.~Liu}
\affiliation{Key Laboratory of Particle and Radiation Imaging (Ministry of Education) and Department of Engineering Physics, Tsinghua University, Beijing 100084}
\author{S.~K.~Liu}
\affiliation{College of Physics, Sichuan University, Chengdu 610065}
\author{Y.~D.~Liu}
\affiliation{College of Nuclear Science and Technology, Beijing Normal University, Beijing 100875}
\author{Y.~Liu}
\affiliation{College of Physics, Sichuan University, Chengdu 610065}
\author{Y.~Y.~Liu}
\affiliation{College of Nuclear Science and Technology, Beijing Normal University, Beijing 100875}
\author{H.~Ma}
\affiliation{Key Laboratory of Particle and Radiation Imaging (Ministry of Education) and Department of Engineering Physics, Tsinghua University, Beijing 100084}
\author{Y.~C.~Mao}
\affiliation{School of Physics, Peking University, Beijing 100871}
\author{Q.~Y.~Nie}
\affiliation{Key Laboratory of Particle and Radiation Imaging (Ministry of Education) and Department of Engineering Physics, Tsinghua University, Beijing 100084}
\author{J.~H.~Ning}
\affiliation{YaLong River Hydropower Development Company, Chengdu 610051}
\author{H.~Pan}
\affiliation{NUCTECH Company, Beijing 100084}
\author{N.~C.~Qi}
\affiliation{YaLong River Hydropower Development Company, Chengdu 610051}
\author{J.~Ren}
\affiliation{Department of Nuclear Physics, China Institute of Atomic Energy, Beijing 102413}
\author{X.~C.~Ruan}
\affiliation{Department of Nuclear Physics, China Institute of Atomic Energy, Beijing 102413}
\author{M.~K.~Singh}
\altaffiliation{Participating as a member of TEXONO Collaboration}
\affiliation{Institute of Physics, Academia Sinica, Taipei 11529}
\affiliation{Department of Physics, Banaras Hindu University, Varanasi 221005}
\author{T.~X.~Sun}
\affiliation{College of Nuclear Science and Technology, Beijing Normal University, Beijing 100875}
\author{C.~J.~Tang}
\affiliation{College of Physics, Sichuan University, Chengdu 610065}
\author{Y.~Tian}
\affiliation{Key Laboratory of Particle and Radiation Imaging (Ministry of Education) and Department of Engineering Physics, Tsinghua University, Beijing 100084}
\author{G.~F.~Wang}
\affiliation{College of Nuclear Science and Technology, Beijing Normal University, Beijing 100875}
\author{J.~Z.~Wang}
\affiliation{Key Laboratory of Particle and Radiation Imaging (Ministry of Education) and Department of Engineering Physics, Tsinghua University, Beijing 100084}
\author{L.~Wang}
\affiliation{Department of  Physics, Beijing Normal University, Beijing 100875}
\author{Q.~Wang}
\affiliation{Key Laboratory of Particle and Radiation Imaging (Ministry of Education) and Department of Engineering Physics, Tsinghua University, Beijing 100084}
\affiliation{Department of Physics, Tsinghua University, Beijing 100084}
\author{Y.~F.~Wang}
\affiliation{Key Laboratory of Particle and Radiation Imaging (Ministry of Education) and Department of Engineering Physics, Tsinghua University, Beijing 100084}
\author{Y.~X.~Wang}
\affiliation{School of Physics, Peking University, Beijing 100871}
\author{H.~T.~Wong}
\altaffiliation{Participating as a member of TEXONO Collaboration}
\affiliation{Institute of Physics, Academia Sinica, Taipei 11529}
\author{S.~Y.~Wu}
\affiliation{YaLong River Hydropower Development Company, Chengdu 610051}
\author{Y.~C.~Wu}
\affiliation{Key Laboratory of Particle and Radiation Imaging (Ministry of Education) and Department of Engineering Physics, Tsinghua University, Beijing 100084}
\author{H.~Y.~Xing}
\affiliation{College of Physics, Sichuan University, Chengdu 610065}
\author{Y.~Xu}
\affiliation{School of Physics, Nankai University, Tianjin 300071}
\author{T.~Xue}
\affiliation{Key Laboratory of Particle and Radiation Imaging (Ministry of Education) and Department of Engineering Physics, Tsinghua University, Beijing 100084}
\author{Y.~L.~Yan}
\affiliation{College of Physics, Sichuan University, Chengdu 610065}
\author{N.~Yi}
\affiliation{Key Laboratory of Particle and Radiation Imaging (Ministry of Education) and Department of Engineering Physics, Tsinghua University, Beijing 100084}
\author{C.~X.~Yu}
\affiliation{School of Physics, Nankai University, Tianjin 300071}
\author{H.~J.~Yu}
\affiliation{NUCTECH Company, Beijing 100084}
\author{J.~F.~Yue}
\affiliation{YaLong River Hydropower Development Company, Chengdu 610051}
\author{M.~Zeng}
\affiliation{Key Laboratory of Particle and Radiation Imaging (Ministry of Education) and Department of Engineering Physics, Tsinghua University, Beijing 100084}
\author{Z.~Zeng}
\affiliation{Key Laboratory of Particle and Radiation Imaging (Ministry of Education) and Department of Engineering Physics, Tsinghua University, Beijing 100084}
\author{B.~T.~Zhang}
\affiliation{Key Laboratory of Particle and Radiation Imaging (Ministry of Education) and Department of Engineering Physics, Tsinghua University, Beijing 100084}
\author{F.~S.~Zhang}
\affiliation{College of Nuclear Science and Technology, Beijing Normal University, Beijing 100875}
\author{L.~Zhang}
\affiliation{College of Physics, Sichuan University, Chengdu 610065}
\author{Z.~H.~Zhang}
\affiliation{Key Laboratory of Particle and Radiation Imaging (Ministry of Education) and Department of Engineering Physics, Tsinghua University, Beijing 100084}
\author{Z.~Y.~Zhang}
\affiliation{Key Laboratory of Particle and Radiation Imaging (Ministry of Education) and Department of Engineering Physics, Tsinghua University, Beijing 100084}
\author{J.~Z.~Zhao}
\affiliation{Key Laboratory of Particle and Radiation Imaging (Ministry of Education) and Department of Engineering Physics, Tsinghua University, Beijing 100084}
\author{K.~K.~Zhao}
\affiliation{College of Physics, Sichuan University, Chengdu 610065}
\author{M.~G.~Zhao}
\affiliation{School of Physics, Nankai University, Tianjin 300071}
\author{J.~F.~Zhou}
\affiliation{YaLong River Hydropower Development Company, Chengdu 610051}
\author{Z.~Y.~Zhou}
\affiliation{Department of Nuclear Physics, China Institute of Atomic Energy, Beijing 102413}
\author{J.~J.~Zhu}
\affiliation{College of Physics, Sichuan University, Chengdu 610065}

\collaboration{CDEX Collaboration}
\noaffiliation

\date{\today}

\begin{abstract}
We report new constraints on light dark matter (DM) boosted by blazars using the 205.4 kg day data from the CDEX-10 experiment located at the China Jinping Underground Laboratory. Two representative blazars, TXS 0506+56 and BL Lacertae are studied. The results derived from TXS 0506+56 exclude DM-nucleon elastic scattering cross sections from $4.6\times 10^{-33}\ \rm cm^2$ to $1\times10^{-26}\ \rm cm^2$ for DM masses between 10 keV and 1 GeV, and the results derived from BL Lacertae exclude DM-nucleon elastic scattering cross sections from $2.4\times 10^{-34}\ \rm cm^2$ to $1\times10^{-26}\ \rm cm^2$ for the same range of DM masses. The constraints correspond to the best sensitivities among solid-state detector experiments in the sub-MeV mass range.
\end{abstract}
\maketitle

\emph{\label{sec1}Introduction.}
---The existence of dark matter (DM, denoted by $\chi$) in the universe is supported by convincing cosmological evidence~\cite{pdg2018,bertone_particle_2005}. Direct detection (DD) experiments such as XENON~\cite{XENON_new}, LUX~\cite{LUX_new}, PandaX~\cite{PandaX_nature}, DarkSide~\cite{darkside}, CRESST~\cite{cresst}, SuperCDMS~\cite{cdmslite}, CoGeNT~\cite{cogent2013} and CDEX~\cite{cdex0,cdex1,cdex12014,cdex12016,cdex1b2018,cdex102018,cdex10_tech,cdex1b_am,cdex10_eft,CDEX_electron,CDEX_neutrino} are dedicated to probing DM-nucleus ($\chi$-N) elastic scattering through spin-independent (SI) and spin-dependent interactions, yet no clear signals have been observed to date.

DD experiments rapidly lose sensitivity toward the sub-GeV mass range, because light DM particles carry insufficient energy to generate nuclear recoil signals that exceed the threshold of the detector. Numerous efforts have been made to extend the DD exclusion region toward the sub-GeV mass range. Several inelastic scattering mechanisms, such as the Migdal effect~\cite{cdexmidgal} and bremsstrahlung emission~\cite{brem_prl}, have been used to improve the sensitivity for low mass DM searches~\cite{cdexmidgal}. Recently, it has been pointed out that DM particles can be boosted to relativistic or near-relativistic momenta by high energy particles in the Galaxy, enabling them to generate sufficient nuclear recoil energy in the detector. Such boosted DM can be produced by cosmic-ray nuclei and electrons (termed the CRDM scenario)~\cite{CRDM_prl,CRDM_neutrino,CRDM_PRD,CRDM_ZYF,CRDM_TJ,PandaX-CRDM,CDEX_CRDM}, astrophysical neutrinos~\cite{vBDM,supernova_v,solar_neutrino,pbh_v}, and high energy electrons inside the Sun~\cite{sun_prl,sun_prd}.

Blazars, a type of Active Galactic Nuclei (AGN), were suggested as a new type of DM booster in recent studies~\cite{BBDM_PRL,BBDM_SK}. The high energy protons and electrons in the blazar jet pointing at the Earth can boost DM particles to very high velocities. The supermassive black hole (BH) in the center of a blazar attracts a dense population of DM particles around it. The blazar-boosting effect can then significantly enhance the DM flux reaching the Earth and improve constraints on the DM elastic scattering cross sections.

CDEX-10, the second phase of the CDEX experiment~\cite{cdex0,cdex1,cdex12014,cdex12016,cdex1b2018,cdex102018,cdex10_tech,cdex10_eft,cdex1b_am,CDEX_CRDM,CDEX_electron,CDEX_exotic,CDEX_neutrino,CDEX_black_hole} aiming at light DM searches, operates a 10-kg $p$-type point contact germanium (PPCGe)~\cite{soma2016} detector array in the China Jinping Underground Laboratory (CJPL), which has a rock overburden of 2400 meters~\cite{cjpl}. The detector array consists of three triple-element PPCGe detector strings that are immersed directly in liquid nitrogen (L$\rm N_2$) for cooling and shielding. The polyethylene room with 1 m thick walls and the 20 cm thick high-purity oxygen-free copper in the L$\rm N_2$ cryostat serve as passive shields against ambient radioactivity. The configuration of the detector system was described in detail previously~\cite{cdex102018,cdex10_tech}. The detector has achieved an energy threshold of 160 eVee (electron equivalent energy) and a background level of about 2 counts/keVee/kg/day (cpkkd)~\cite{cdex_darkphoton}. The dataset has been analyzed within the CRDM scenario and has expanded the constraints to $\mathcal{O}(10^{-30})\ \rm cm^2$ in the sub-GeV region~\cite{CDEX_CRDM}. 

In this letter, we reanalyzed the 205.4 kg day data set from the CDEX-10 experiment~\cite{cdex_darkphoton} within the blazar-boosted DM (BBDM) scenario to set constraints on the $\chi$-nucleon SI interactions, the cross sections for which we denote by $\sigma_{\chi N}$. The blazars considered in this analysis are TXS 0506+56, which may be the source of the high energy neutrinos detected by the IceCube Observatory~\cite{IceCube1,IceCube2}, and a typical BL Lac object BL Lacertae~\cite{Abdo_2010}. We utilize the CJPL\_ESS simulation package to evaluate the Earth shielding effect~\cite{earthshielding,earthshielding1,earthshielding2,earthshielding3,earthshielding4,cjpless}, considering the influence of the rock overburden during the DM transportation to the underground laboratory to place an upper bound on the exclusion region. 

\emph{\label{sec2}Blazar jet spectrum.} 
---Particles inside a blazar are accelerated into two back-to-back jets, one of which is closely aligned with our line of sight (LOS)~\cite{blazar_los}. The spectral energy distribution (SED) of non-thermal photons from a blazar has two peaks, one in the infrared band and the other in the $\gamma$ ray band~\cite{blazar_sed}. The low energy peak is thought to be generated by synchrotron emission from electrons, while the origin of the high energy peak remains unclear. According to the leptonic model, blazar electrons may be responsible for it, while the pure hadronic and hybrid lepto-hadronic models claim that protons may also contribute to the SED~\cite{BL_parameter}. The SED models for different blazars can be derived from the data measured with the Fermi-Large Area Telescope (Fermi-LAT) and the Air Cherenkov Telescope~\cite{fermi1,fermi2}. Due to the high time variability and population variability of blazars, the model parameters rely significantly on the source considered and the observation time.

In this work, we adopt the lepto-hadronic model, which is in good agreement with the SED of TXS 0506+56~\cite{txs_lepto,txs_lepto2}. For BL Lacertae, we adopt the hadronic model~\cite{BL_parameter}.

The jet geometry can be simplified into a ''blob'', in which the particles (electrons and protons) move isotropically with a power-law energy distribution. The blob moves along the jet axis with speed $\beta_B$ in the observer's frame along the jet axis, which is inclined by an angle $\theta_{\rm LOS}$ with respect to the observer's line of sight. The corresponding Lorentz factor is $\Gamma_B=1/\sqrt{1-\beta_B^2}$, while $\mathcal{D}=[\Gamma_B(1-\rm cos(\theta_{LOS}))]^{-1}$ is the Doppler factor. Two common assumptions used in the model fitting is $\mathcal{D}=2\Gamma_B$ and $\Gamma_B$, corresponding to the case $\theta_{\rm LOS}=0$ and $\theta_{\rm LOS}=1/\mathcal{D}$ ($\mathcal{D}>>1$), respectively.

Following the method described in Ref.~\cite{BBDM_PRL}, the proton spectrum of the blazar jet in the observer's frame can be expressed as:
\begin{equation}
\frac{d\Gamma_p}{dT_pd\Omega}=\frac{c_p}{4\pi}(1+\frac{T_p}{m_p})^{-\alpha_p}\frac{\beta_p(1-\beta_p\beta_B\mu)^{-\alpha_p}\Gamma_B^{-\alpha_p}}{\sqrt{(1-\beta_p\beta_B\mu)^2-(1-\beta_p)(1-\beta_B)}},
\label{eq_flux}
\end{equation}
where $T_p$ is the kinetic energy of a blazar proton in the observer's frame, $m_p$ is the proton mass, $\alpha_p$ is the power index, $\beta_p$ is the proton speed, $\mu$ is the cosine of the angle between the LOS and the jet axis, and $c_p$ is the normalization factor that can be computed from the overall luminosity $L_p$. The model parameters for the blazar TXS 0506+56 and for BL Lacertae are listed in Table~\ref{tab:table1}, together with the redshift $z$, luminosity distance $d_L$ and mass $M_{\rm BH}$ of the central BH. The data labeled with a star (*) correspond to the mean value calculated from the range given in  the second column of Table 1 of Ref.~{\cite{txs_lepto2}.} The high energy protons and electrons can both serve as DM boosters. This analysis only considers the protons in the blazars as our main focus is $\chi$-N scattering. The blazar electron boosing effect is left to future analysis.

\begin{table}[!htbp]
\caption{\label{tab:table1}
The Lepto-Hadronic (Hadronic) model parameters for the blazar TXS 0506+56 (BL Lacertae)~\cite{txs_lepto,txs_lepto2,BL_parameter}. The data labeled with a star (*) correspond to the mean values calculated from the range given in the second column of Table 1 of Ref.~\cite{txs_lepto2}.}
\begin{ruledtabular}
\begin{tabular}{lccccccccccccccc}
  \multicolumn{1}{c}{Parameters}
         &   \multicolumn{1}{c}{TXS 0506+56}     & \multicolumn{1}{c}{BL Lacertae} 
          \\
\hline
\multirow{1}{*}{$z$} & 0.337 & 0.069     \\
\multirow{1}{*}{$d_L$ (Mpc)} & 1835.4 & 322.7  \\
\multirow{1}{*}{$M_{\rm BH}$ ($M_{\odot}$)} & $3.09\times 10^8$ & $8.65\times 10^7$  \\
\multirow{1}{*}{$\mathcal{D}$} & $40^*$ & 15  \\
\multirow{1}{*}{$\Gamma_B$} & 20 & 15  \\
\multirow{1}{*}{$\theta_{\rm LOS}(^\circ)$} & 0 & 3.82  \\
\multirow{1}{*}{$\alpha_{p}$} & $2.0^*$ & 2.4  \\
\multirow{1}{*}{$\gamma'_{p,min}$} & 1.0 & 1.0  \\
\multirow{1}{*}{$\gamma'_{p,max}$} & $5.5\times 10^{7*}$ & $1.9\times 10^9$  \\
\multirow{1}{*}{$L_{p}$(erg/s)} & $2.55\times 10^{48*}$ & $9.8\times 10^{48}$  \\
\end{tabular}
\end{ruledtabular}
\end{table}

\emph{\label{sec31}The dark matter density profile.}
---The adiabatic growth of a BH in the central region of a DM halo can concentrate the DM density significantly, creating a very dense spike~\cite{DM_profile}. If the DM density initially follows a power law $\rho(r)=r^{-\gamma}$, the profile after evolution can be expressed as
\begin{equation}
    \rho_{\rm DM} =\frac{\rho'(r)\rho_{\rm core}}{\rho'(r)+\rho_{\rm core}},
\end{equation}
where
\begin{equation}
    \rho'(r)=r^{-\frac{9-2\gamma}{4-\gamma}}
\end{equation}
is the spike distribution. The correction to $\rho_{\rm core}$ due to DM annihilation $\rho_{\rm core}$ is $\rho_{\rm core}=m_{\chi}/\left\langle\sigma v\right\rangle_0t_{\rm BH}$, where $\left\langle\sigma v\right\rangle_0$ is the DM annihilation cross section multiplied by the relative velocity, and $t_{\rm BH}$ is the BH lifetime. In this analysis, we adopt $\gamma =1$. The normalization condition for $\rho'(r)$ is given by
\begin{equation}
    \int_{4R_S}^{10^5R_S}4\pi r^2\rho'(r)dr \approx M_{\rm BH}
\end{equation}
where $R_S$ is the Schwarzschild radius of the central BH. Two benchmark points (BMP, denoted BMP1 and BMP2) are discussed in previous works:

BMP1: $\left\langle\sigma v\right\rangle_0=0$, so that $\rho_{\rm core}\rightarrow \infty$ and $\rho_{\rm DM}=\rho'$;

BMP2: $\left\langle\sigma v\right\rangle_0=10^{-28}{\rm cm^3}{\rm s}^{-1}$ and $t_{\rm BH}=10^9 $ yr.

\begin{figure}[!htbp]
\includegraphics[width=\linewidth]{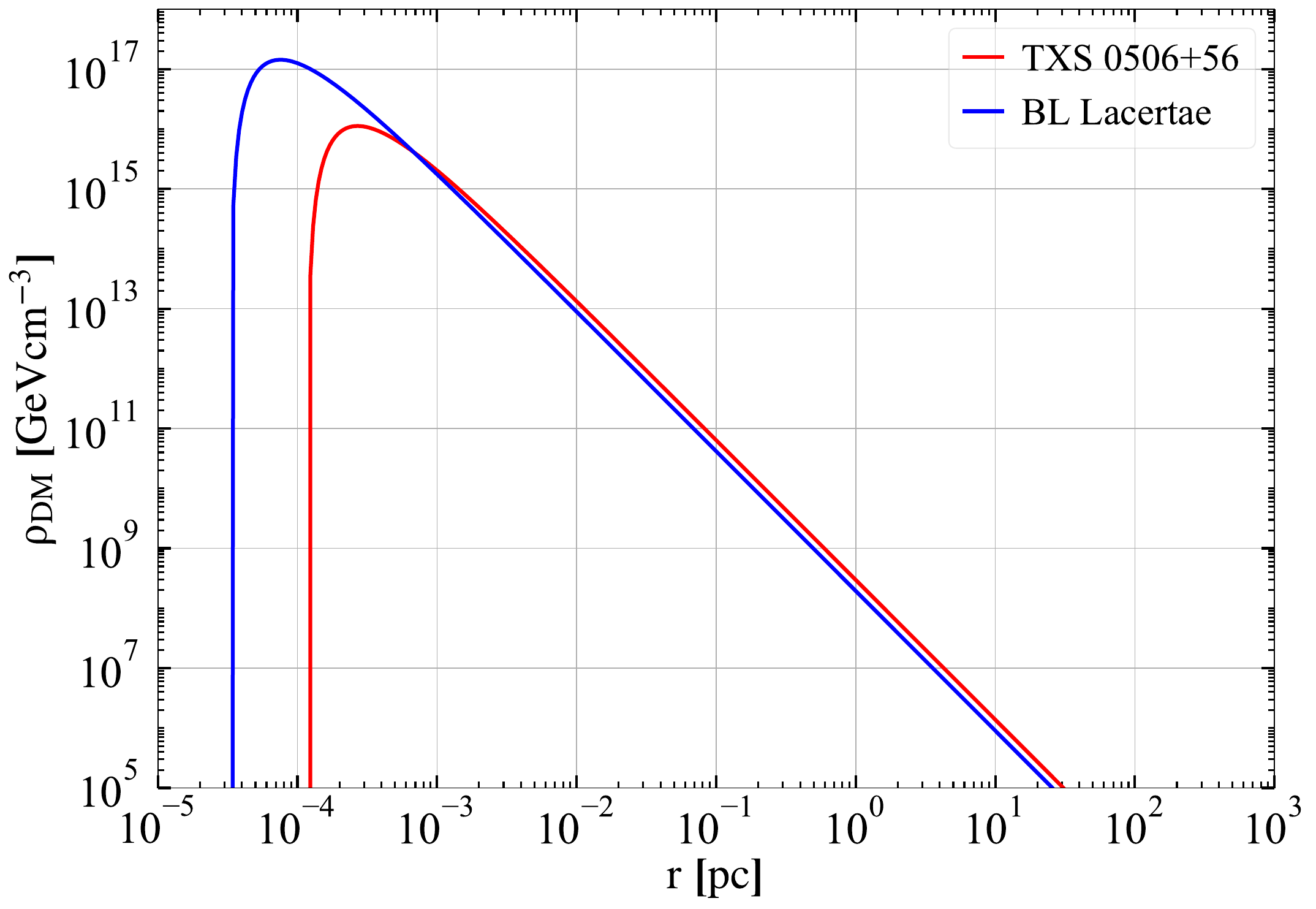}
\caption{
The dark matter density distributions $\rho_{DM}$ for TXS 0506+56 (red) and BL Lacertae (blue) with $m_{\chi}=1$ MeV.
}
\label{fig::rho}
\end{figure}

The BMP1 world be more appropriate for asymmetric DM models~\cite{BBDM_PRL} and only the BMP1 case is analyzed in the following analysis. The dark matter density profiles corresponding to BMP1 for the blazar TXS 0506+56 and for BL Lacertae are demonstrated in Fig.~\ref{fig::rho}. The accumulated DM mass, $\Sigma_{\rm DM}$, used in calculating the BBDM flux is given by:
\begin{equation}
    \Sigma_{\rm DM}(r)=\int_{4R_S}^{r}\rho_{\rm DM}(r')dr'.
\end{equation}

As shown in Fig.~\ref{fig::rho}, the dark matter density profiles drop drastically as $r\rightarrow \infty$. The value of $\Sigma_{\rm DM}$ tends to saturate when $r>10 $ kpc, so we set the upper limit of the integration to $10^5 R_S$, which has little impact on the final results.

\emph{\label{sec32}The blazar boosted dark matter flux.}
---The BBDM flux reaching the Earth that is induced by particle proton can be expressed as:
\begin{equation}
    \frac{d\Phi_{\chi}}{dT_{\chi}}=\frac{\Sigma^{tot}_{\rm DM}}{2\pi m_{\chi}d^2_L}\sigma_{\chi p}\int_0^{2\pi}d\phi_S\int_{T_p^{min}(T_{\chi})}^{T_p^{max}(T_{\chi})}\frac{dT_p}{T_{\chi}^{max}(T_p)}\frac{d\Gamma_p}{dT_p d\Omega}
\end{equation}
where $\phi_S$ is the azimuth angle. The maximal kinetic energy that a DM particle can have after one collision is 
\begin{equation}
T_{\chi}^{max}(T_p)=\frac{T_p^2+2m_pT_p}{T_p+(m_p+m_{\chi})^2/(2m_{\chi})} .
\label{Tmax}
\end{equation}

The minimal energy $T_p^{min}$ required to produce the DM kinetic energy $T_{\chi}$ can be obtained by inverting Eq.~\ref{Tmax}.  As the blazar proton flux drops quickly in the high energy region, the upper bound of the integration has little influence on the results, so we set $T_p^{max}=10^7 $ GeV.

\begin{figure}[!htbp]
\includegraphics[width=\linewidth]{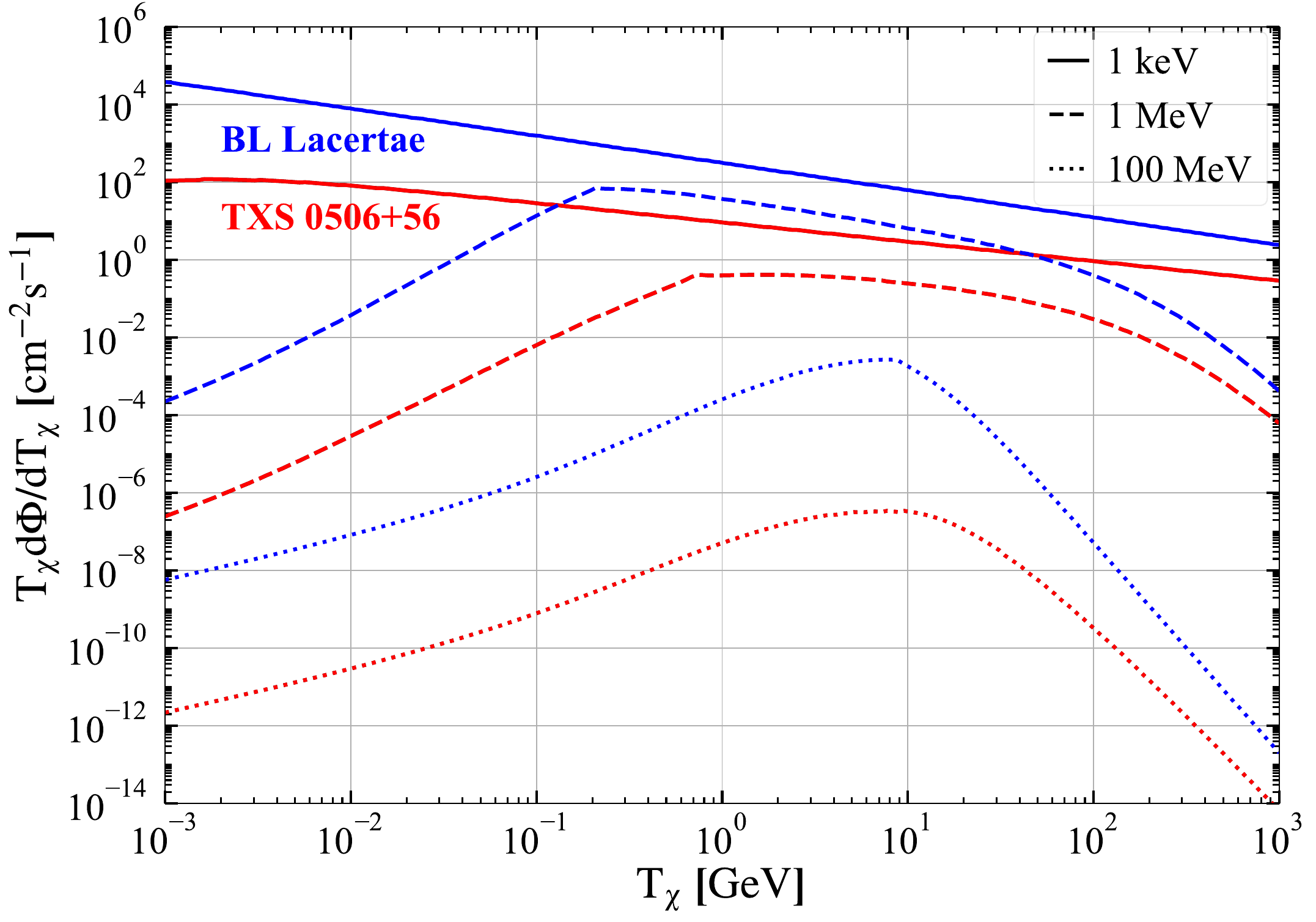}

\caption{The expected dark matter flux boosted by protons for TXS 0506+56 (red) and BL Lacertae (blue). Different line styles correspond to different dark matter masses.}
\label{fig::bbdm}
\end{figure}

The $\chi-p$ differential scattering cross section is given by:
\begin{equation} \label{sigma}
 \frac{d\sigma_{\chi p}}{dT_{\chi}}=\frac{\sigma_{\chi N}}{T_{\chi}^{max}}A_p^2(\frac{\mu_{\chi p}}{\mu_{\chi N}})^2G_p(Q^2),    
\end{equation}
where $\sigma_{\chi N}$ is the zero-momentum transfer DM-nucleon cross section, $A_i$ is the mass number of blazar component $i$, $\mu_{\chi i}$ is the DM-nucleus reduced mass, $\mu_{\chi N}$ is the DM-nucleon reduced mass, and $G$ is the form factor, which is related to the momentum transfer, $Q=\sqrt{2m_{\chi}T_{\chi}}$. For protons, we adopt the dipole form factor $G_i(Q^2)=1/(1+Q^2/\Lambda_i^2)^2$~\cite{dipole}, where $\Lambda_p\simeq770$ MeV\cite{p_he_form}. The BBDM fluxes calculated for different DM masses for blazar TXS 0506+56 and BL Lacertae are shown in Fig.~\ref{fig::bbdm}.

\emph{\label{sec4}Earth attenuation and recoil spectrum.}
---The trajectories and kinetic energies of the DM particles will be altered after traveling through a few kilometers of rock before reaching the underground laboratory. In this analysis, we used the Monte-Carlo simulation package CJPL\_ESS~\cite{cjpless} to evaluate the Earth shielding effect. The CJPL\_ESS software includes the detailed geometric model and the rock compositions of the Jinping Mountain. This simulation package also considers angular deflections and the impact of the form factor in the scattering process between the DM particles and the compositions of the rock overburden, which can significantly enhance the DM flux arriving at the underground laboratory.

The differential event rate of $\chi$-N elastic scattering in the detectors is calculated from:
\begin{equation}
\frac{dR}{dE_R}=N_TA^2 (\frac{\mu_{\chi A}}{\mu_{\chi N}})^2\int_{T^{min}_{\chi}}\frac{\sigma_{\chi N}}{E_{R}^{max}} G_A^2(Q^2)\frac{d\Phi_{\chi}}{dT^z_{\chi}}dT^z_{\chi},
\end{equation}
where $E_R$ is the nuclear recoil energy, $N_T$ is the number of target nuclei per unit detector mass, $A$ is the mass number of Ge nucleus and $\mu_{\chi A}$ is the DM-Ge nucleus reduced mass. We obtained the value of $E_{R}^{max}$ from Eq.~\ref{Tmax} by replacing $p\rightarrow\chi$ and $\chi \rightarrow N$. Inverting the expression of $E_{R}^{max}$ gives $T^{min}_{\chi}$. The $G_A(Q^2)$ is the nuclear form factor, for which the Helm form factor~\cite{formfactor2} is used.

\begin{figure}[!htbp]

\centering\includegraphics[width=\linewidth]{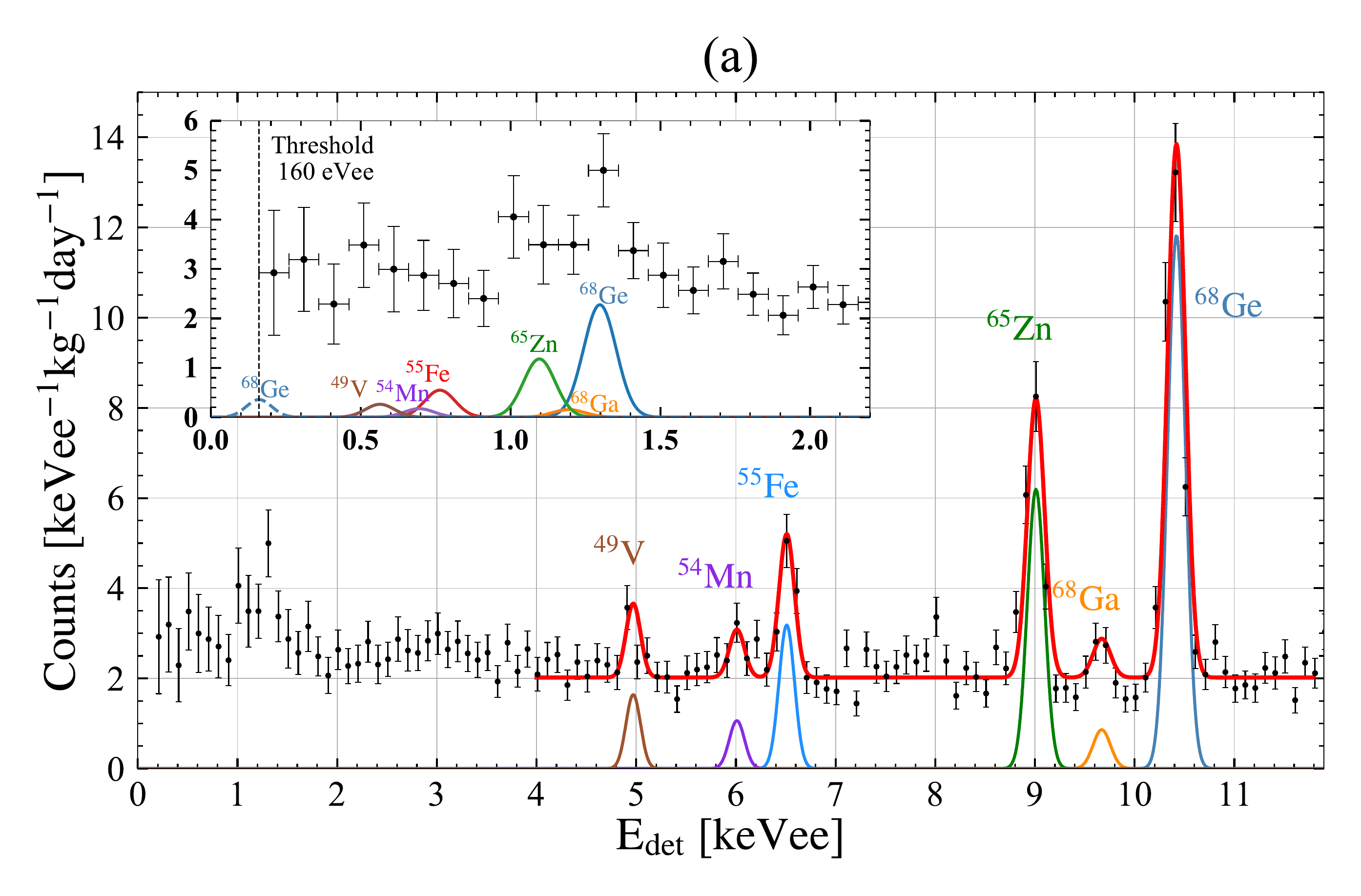}
\includegraphics[width=\linewidth]{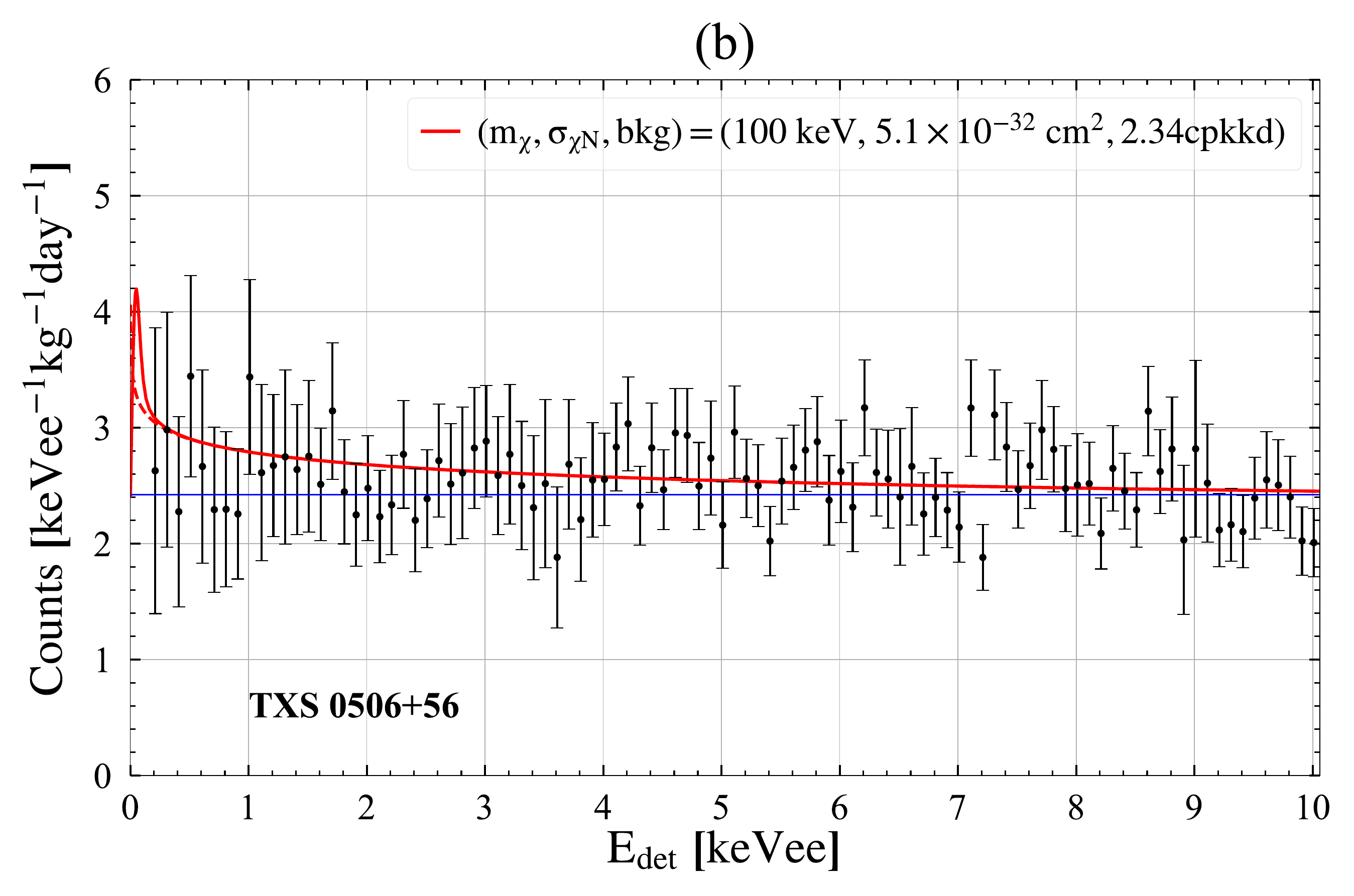}
\caption{(a) The measured energy spectrum with error bars from the CDEX-10 205.4 kg day dataset, together with the characteristic K-shell x-ray peaks from internal cosmogenic radionuclides. Both the best fit curve of the measured energy spectrum in the range 4-11.8 keV (red line) and the contributions of these radionuclides derived from the best fit are superimposed. The contributions of the L- and M-shell x-ray peaks derived from the corresponding K-shell line intensities are displayed in the inset. (b) The residual spectrum with error bars from 0 keV to 10.06 keV, with the contributions from the L- and M-shell x-rays subtracted. The BBDM spectrum boosted by TXS 0506+56 corresponds to the lower limits of the exclusion region with a flat background are shown in (b). The solid and dashed lines correspond to spectra with and without considering the energy resolution.}
\label{fig::spectrum}
\end{figure}

The observed total deposit energy $E_{det}$ in one germanium semiconductor detector is different from the real nuclear recoil energy $E_R$. It must be corrected by the quenching factor $Q_{nr}$: $E_{det}=Q_{nr}E_R$. The quenching factor in Ge is calculated using the Lindhard formula~\cite{lindhard} ($\kappa = 0.16$, the usual value from the literature and matches recent low-energy measurements~\cite{quenching2022}), with a 10\% systematic error adopted in this analysis.

The spectra in the detector resulting from the BBDM fluxes boosted by the blazar TXS 0506+56 are shown in Fig.~\ref{fig::spectrum}, together with a flat background. The solid and dashed lines correspond to the spectra with and without considering the energy resolution.

\emph{\label{sec5}Exclusion results.}
---The data used in this BBDM analysis are from the CDEX-10 experiment with a total exposure of 205.4 kg day~\cite{cdex_darkphoton}. The data analysis follows the procedures described in our earlier works, including energy calibration, physics event selections, bulk-surface event discrimination, and various efficiency corrections~\cite{cdex1b2018,cdex10_tech,cdex102018,cdex1b_am,cdex_darkphoton}. The physics analysis threshold is 160 eVee where the combined efficiency (including the trigger efficiency and the efficiency for pulse shape discrimination) is 4.5\%. The characteristic K-shell X-ray peaks from internal cosmogenic radionuclides like $\rm ^{68}Ge$, $\rm ^{68}Ga$, $\rm ^{65}Zn$, $\rm ^{55}Fe$, $\rm ^{54}Mn$ and $\rm ^{49}V$ are fitted and subtracted from the spectrum, as shown in Fig.~\ref{fig::spectrum}(a). The intensities of L- and M-shell x-ray peaks are derived from the K-shell x-ray peaks. The standard deviation of the energy resolution is 35.8 + 16.6$\times E^{\frac {1}{2}}$ (eV), where $E$ is expressed in keV~\cite{cjpless}.

A minimal-$\chi^2$ method is applied to derive the exclusion region, following the treatment in our previous works~\cite{cdex12014,cdex102018,cdex_darkphoton}. The $\chi^2$ value at certain DM mass $m_{\chi}$ and cross section $\sigma_{\chi N}$ is defined by:
\begin{equation}
    \chi^2(m_{\chi},\sigma_{\chi N}) = \sum_{i=0}^{N_j}\frac{[n_i-S_i(m_{\chi},\sigma_{\chi N})-B_i]^2}{\sigma^2_{stat, i}+\sigma^2_{syst, i}},
\end{equation}
where $n_i$ is the measured event rate corresponding to the $i_{th}$ energy bin, and $S_i(m_{\chi},\sigma_{\chi N})$ is the expected rate. Both statistical (stat) and systematical (syst) uncertainties are included, denoted by $\sigma_{stat}$ and $\sigma_{syst}$, respectively. The background contribution at the $i_{th}$ energy bin is represented by $B_i$, which we assume to be a flat distribution; i.e., $B_i=B$. The best estimator of $\sigma_{\chi N}$ at a given DM mass is obtained by minimizing the $\chi^2$ value, from which upper limits at the 90\% confidence level (C.L.) are derived. The energy range of the $\chi^2$ fit is from 0.16 keV to 10.06 keV.

\begin{figure}[!htbp]
\includegraphics[width=\linewidth]{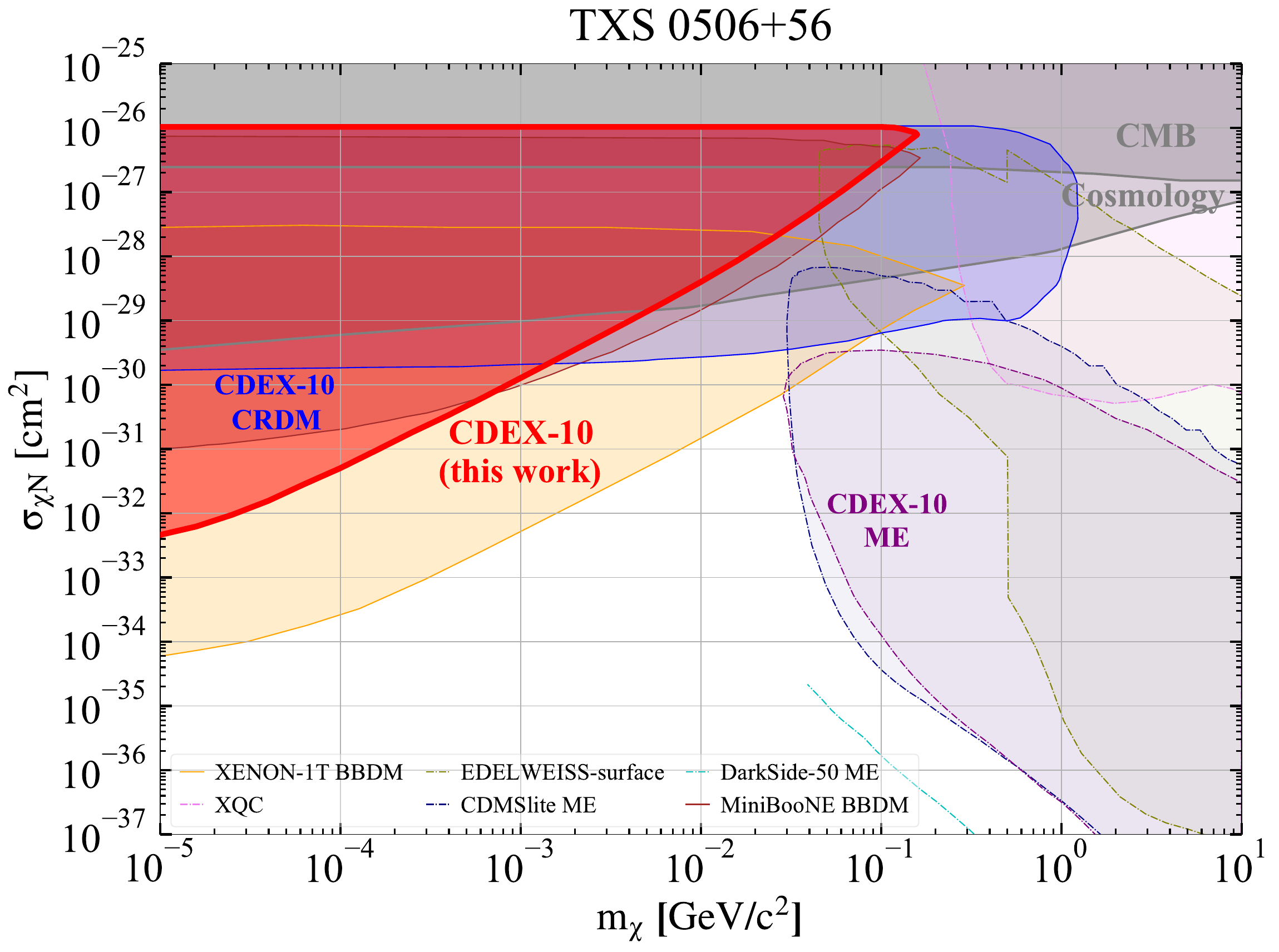}
\includegraphics[width=\linewidth]{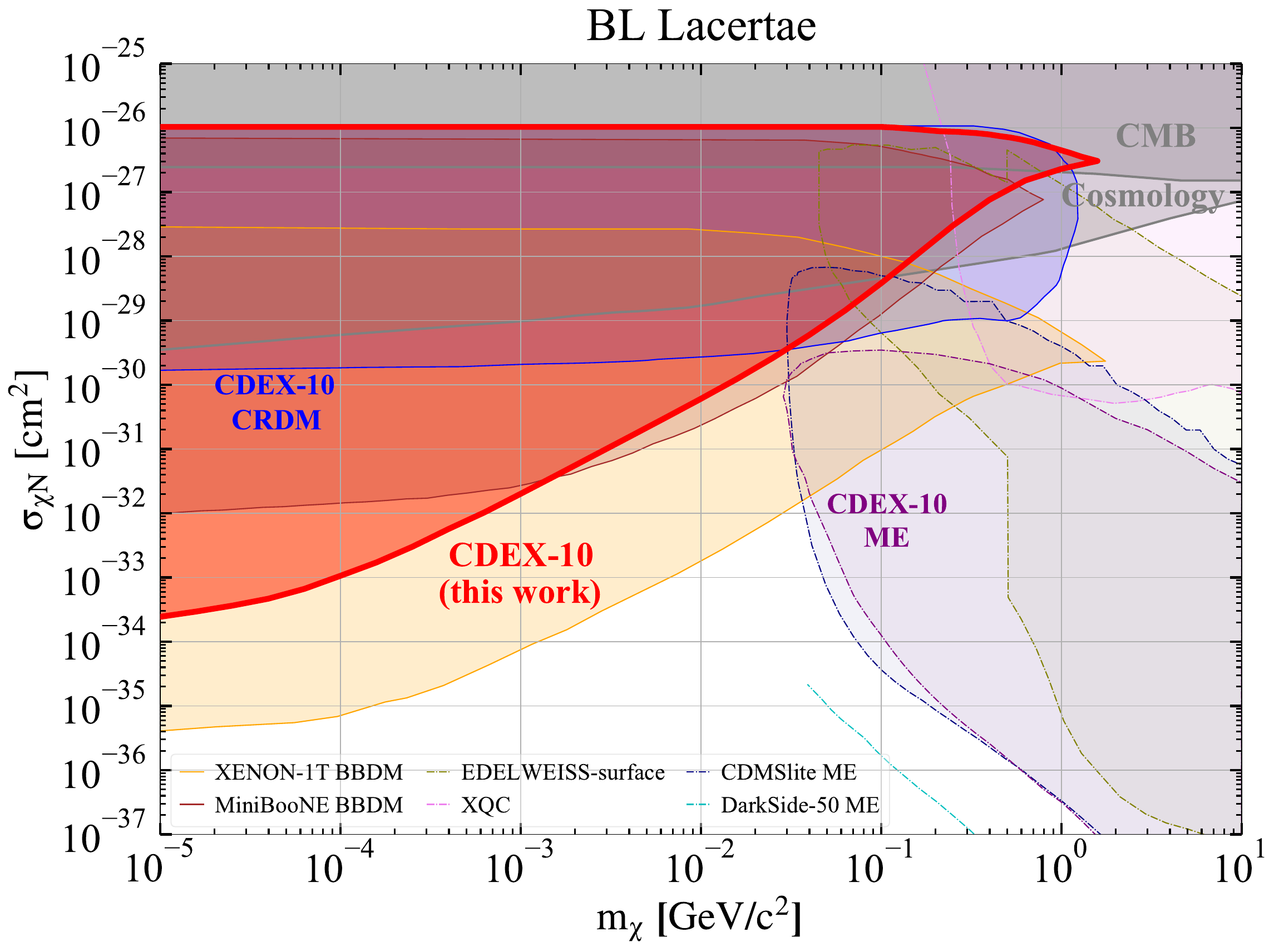}
\caption{
The exclusion regions derived from the CDEX-10 experiment are outlined with solid red contours, with the blazar boosting mechanism adopted. The constraints derived from the blazar TXS 0506+56 and BL Lacertae in the BMP1 case are shown in the upper and lower panels, respectively. The limits obtained from a phenomenological analysis using the data from XENON-1T are also demonstrated~\cite{BBDM_PRL}. The blue regions were obtained from the analysis of CDEX-10 data under the CRDM scenario~\cite{CDEX_CRDM}. Other published limits with standard halo model (SHM) DM assumptions from the CDEX-10 Migdal effect (ME) analysis~\cite{cjpless}, CDMSlite ME analysis~\cite{SuperCDMS:2023ME}, DarkSide-50 ME analysis~\cite{DarkSide:2022ME}, EDELWEISS-Surface~\cite{edelweiss}, and X-ray Quantum Calorimeter experiment (XQC)~\cite{XQC1,XQC2} are superimposed. Constraints provided by the cosmic microwave background (CMB)~\cite{cmb} and the large scale structure of the universe (Cosmology)~\cite{cosmo11} are also shown.
}
\label{fig::region}
\end{figure}

The exclusion regions for BBDM with $90\%$ C.L. are derived using the minimal-$\chi^2$ method and shown in Fig.~\ref{fig::region}. The limits set by former phenomenological studies using XENON-1T data and the blazar boosting effect are also shown. The exclusion region from CDEX-10 using CR acceleration (CRDM)~\cite{CDEX_CRDM}, and the published limits under the SHM scenario from CDEX-10 Migdal effect (ME) analysis~\cite{cjpless},CDMSlite ME analysis~\cite{SuperCDMS:2023ME}, DarkSide-50 ME analysis~\cite{DarkSide:2022ME}, EDELWEISS-Surface~\cite{edelweiss}, and X-ray Quantum Calorimeter experiment (XQC)~\cite{XQC1,XQC2} are also superimposed. The result derived from the blazar TXS 0506+56 excludes DM-nucleon elastic scattering cross sections from $4.6\times 10^{-33}\ \rm cm^2$ to $1\times10^{-26}\ \rm cm^2$ for DM masses from 10 keV to 1 GeV, while the result derived from BL Lacertae excludes DM-nucleon elastic scattering cross sections from $2.4\times 10^{-34}\ \rm cm^2$ to $1\times10^{-26}\ \rm cm^2$ for DM mass from 10 keV to 1 GeV.

\emph{\label{sec6} Conclusions and discussion.}
---We report new limits on the blazar boosted dark matter using the CDEX-10 experiments data. The exclusion results are improved by approximately two orders of magnitude at a DM mass of 10 keV compared to the results derived from the CR acceleration (CRDM)~\cite{CDEX_CRDM}. The constraints correspond to the best sensitivities among experiments using solid-state detectors in DM mass range 10 keV-0.8 MeV for TXS 0506+56 BMP1 and 10 keV-1.5 MeV for BL Lacertae BMP1. The high exposure and low background of the XENON1T experiment enable it to provide more sensitive constraints~\cite{BBDM_PRL}.

The present analysis considers only the DM-nucleon SI elastic scattering process. The population of high-speed electrons in blazars can also boost the dark matter, giving rise to novel DM-electron cross section constraints~\cite{BBDM_SK}. The scattering process between high velocity DM particles and the electrons in semiconductor detectors is rather complicated. With the development of computing techniques, novel DM-e scattering constraints will be derived from the CDEX experiment~\cite{electron, CDEX_electron}. The cross sections given by Eq.~\ref{sigma} are energy independent if form factor corrections are included. The results can be improved further if an energy-dependent cross section is considered~\cite{energy_dependent,Bardhan:2022bdg}.

In addition, we have considered only the two blazars TXS 0506+56 and BL Lacertae  in the present BBDM analyses. Multiple other blazars are also of interest to astronomical researchers and have benn proposed as sources of boosted DM~\cite{PKS1502+106,Franckowiak:2020qrq,Gasparyan:2021oad}. The fluxes of the protons and electrons and the distances between the considered blazar and the Earth are the main factors that affect the BBDM fluxes. Blazars with larger proton and electron fluxes and smaller distances to the Earth can yield more stringent limits than the present results derived from two blazars. Blazar fluxes also vary significantly over time, and this can be used to study modulation of the BBDM spectrum~\cite{Gasparyan:2021oad}. 

This work was supported by the National Key Research and Development Program of China (Grants No. 2017YFA0402200, No. 2022YFA1605000) and the National Natural Science Foundation of China (Grants No. 12322511, No. 12175112, No. 12005111, and No. 11725522). We acknowledge the Center of High performance computing, Tsinghua University for providing the facility support. We would like to thank CJPL and its staff for hosting and supporting the CDEX project. CJPL is jointly operated by Tsinghua University and Yalong River Hydropower Development Company.

\bibliography{BBDM}

\end{document}